\documentclass[12pt,preprint]{aastex}
\begin{document}

\title{Discovery of a tight correlation for gamma ray burst afterglows with `canonical' light curves}

\author{Maria Giovanna Dainotti\altaffilmark{1}, Richard Willingale\altaffilmark{2}, Salvatore Capozziello\altaffilmark{3,4}, \\ Vincenzo Fabrizio Cardone\altaffilmark{4,5}, Micha{\l} Ostrowski\altaffilmark{1}}

\altaffiltext{1}{Obserwatorium Astronomiczne, Uniwersytet Jagiello\'nski, ul. Orla 171, 31-501 Krak{\'o}w, Poland E-mails: mariagiovannadainotti@yahoo.it, mio@oa.uj.edu.pl}

\altaffiltext{2}{Department of Physics \& Astronomy, University of Leicester, Road Leicester LE1 7RH, United Kingdom E-mail: rw@star.le.ac.uk}

\altaffiltext{3}{Dipartimento di Scienze Fisiche, Universit\`{a} di Napoli "Federico II", Complesso
Universitario di Monte Sant'Angelo, Edificio N, via Cinthia, 80126 - Napoli, Italy E-mails: capozziello@na.infn.it, cardone@to.infn.it} 

\altaffiltext{4}{I.N.F.N., Sez. di Napoli, Complesso Universitario di Monte
Sant' Angelo, Edificio G, via Cinthia, 80126 - Napoli, Italy E-mail: capozziello@na.infn.it}

\altaffiltext{5}{Dipartimento di Fisica Generale, Universit\`{a} di Torino, and I.N.F.N. 
- Sez. di Torino, Via Pietro Giuria 1, 10125 - Torino, Italy E-mail: cardone@to.infn.it}

\date{\today}
\begin{abstract}
Gamma Ray Bursts (GRB) observed up to redshifts $z>8$ are fascinating objects to study due to their still unexplained relativistic outburst mechanisms and a possible use to test cosmological models. Our analysis of 77 GRB afterglows with known redshifts revealed a physical subsample of long GRBs with canonical {\it plateau breaking to power-law} light curves with a significant {\it luminosity $L^*_X$ - break time $T^*_a$} correlation in the GRB rest frame. This subsample forms approximately the {\it upper envelope} of the studied distribution. We have also found a similar relation for a small sample of GRB afterglows that belong to the intermediate class (IC) between the short and the long ones. It proves that within the full sample of afterglows there exist physical subclasses revealed here by tight correlations of their afterglow properties. The afterglows with regular (`canonical') light curves  obey not only a mentioned tight physical scaling, but -- for a given $T^*_a$ -- the more regular progenitor explosions lead to preferentially brighter afterglows. 
\end{abstract}

\keywords{gamma-rays bursts: general - radiation mechanisms: nonthermal}

\maketitle

\section{Introduction}
The detection of GRBs up to high redshifts (z=8.2; \citet{Salvaterra2009}, \citet{Tanvir}), larger than Supernovae Ia (SNeIa) ($z_{max} = 1.77$; \citet{Riess07}), makes these objects appealing for possible use in cosmology. The problem is that GRBs seem not to be standard candles, with their energetics spanning over seven orders of magnitude. Anyway, several GRB luminosity indicators \citep{Amati08,FRR00,N000,G04,liza05,LZ06,Ghirlanda06} and their use to constrain cosmological parameters \citep{fir06,liza05,LWZ09,Qi09,Izzo09} have been proposed till now. Furthermore \cite{Cardone09} have derived an updated GRBs Hubble diagram using the $\log L^*_X$--$\log T^*_a$ (`LT') \footnote{We use the index `$*$' to indicate quantities measured in the GRB rest frame in which $L^*_X \equiv L^*_X(T^*_a)$ is an isotropic X-ray luminosity in the time $T^*_a$, the transition time separating the afterglow plateau and the power-law decay phases \citep{DCC}.} correlation with five other 2D GRBs correlations used by \cite{S07}. However, the problem of large data scatters in the considered luminosity relations \citep{Butler2009,Yu09} and a possible impact of detector thresholds on cosmological standard candles \citep{Shahmoradi09} have been discussed controversially \citep{Cabrera2007}. Among these attempts, \citet{DCC} have proposed a way to standardize GRBs as distance indicator with the discovery of the LT anti-correlation, confirmed by \citet{Ghisellini2009} and \citet{Yamazaki09}. The fitted power-law relation is $\log L^*_X =  \log a + b \cdot \log T^*_a$; the constants $a$ and $b$ are determined using the \citet{Dago05} method. 

In this Letter we study the LT correlation using the extended GRB data set and we demonstrate the existence of a physical LT scaling for 'canonical' light curves in the GRB rest frame. The regular lightcurve afterglows conform rather tightly to this scaling, while the more irregular ones are systematically fainter. A similar correlation is revealed for a subsample of GRB afterglows that belong to the intermediate class (IC). Revealing these physical correlations can help the (still unclear) interpretation of the physical mechanisms responsible for the GRB X-Ray afterglow emission and can infer important information about the nature of the emitting source.  

\section{Data selection and analysis}

We have analyzed a sample of all afterglows with known redshifts detected by Swift from January 2005 up to April 2009, for which the lightcurves include early XRT data and therefore can be fitted by a Willingale's phenomenological model \citep{W07}. The redshifts $z$ are taken from the Greiner's web page http://www.mpe.mpg.de/$\sim$jcg/grb.html~. We have compared these redshifts with the values reported by \citet{Butler2007} and we find that they agree well apart from two cases of GRB 050801 and 060814, but Butler (private communication) suggested that we should use the Greiner redshifts for those two cases. For original references providing the redshift data see \citep{Butler2007,Butler2010}. Our data analysis, including derivation of $T^*_a$ and $L^*_X$ (in units of [s] and [erg/s], respectively) for each afterglow, follows \citet{DCC} and \citet{W07}. The source rest frame luminosity in the Swift XRT bandpass, $(E_{min}, E_{max})=(0.3,10)$ keV, is computed from the equation:

\begin{equation}
L^*_X (E_{min},E_{max},t)= 4 \pi D_L^2(z) \, F_X (E_{min},E_{max},t) \cdot K
\label{eq: lx}
\end{equation}
where $D_L(z)$ is the GRB luminosity distance for the redshift $z$, computed assuming a flat $\Lambda$CDM cosmological model with $\Omega_M = 0.291$ and $h = 0.697$, $F_X$ is the measured X-ray energy flux (in [${\rm erg/cm^2/s}$]) and  $K$ is the K-correction for cosmic expansion.  Using the \citet{B01} expression for the $K$-correction and with $f(t)$ being the Swift XRT lightcurve we have the relation\,:

\begin{equation}
K F_X(E_{min},E_{max},t) = f(t) \ {\times} \
\frac{\int_{E_{min}/(1 + z)}^{E_{max}/(1 + z)}{E \Phi(E)
dE}}{\int_{E_{min}}^{E_{max}}{E \Phi(E) dE}}
\label{eq: fx}
\end{equation}
where $\Phi(E)$ is a usual differential photon spectrum assumed to be $ \propto E^{-\gamma_{a}}=E^{-(\beta_a+1)}$, where $\gamma_a$ and $\beta_a$ are the photon index and the spectral index, respectively. \citet{W07} proposed a functional form for $f(t)$:

\begin{equation}
f(t) = f_p(t) + f_a(t)
\label{eq: fluxtot}
\end{equation}
where the first term accounts for the prompt (the index ``p'') $\gamma$\,-\,ray emission and the initial X\,-\,ray decay, while the second one describes the afterglow (the index ``a''). Both components are modeled with the same functional form\,:

\begin{equation}
f(t) = \left \{
\begin{array}{ll}
\displaystyle{F_c \exp{\left ( \alpha_c - \frac{t \alpha_c}{T_c} \right )} \exp{\left (
- \frac{t_c}{t} \right )}} & {\rm for} \ \ t < T_c \\
~ & ~ \\
\displaystyle{F_c \left ( \frac{t}{T_c} \right )^{-\alpha_c}
\exp{\left ( - \frac{t_c}{t} \right )}} & {\rm for} \ \ t \ge T_c \\
\end{array}
\right .
\label{eq: fc}
\end{equation}

\noindent
where $c$ = $p$ or $a$. The transition from the exponential to the power law occurs at the point $(T_{c},F_{c})$ where the two functional sections have the same value and gradient. The parameter $\alpha_{c}$ is the temporal power law decay index  and the time $t_{c}$ is the the initial rise time scale (for further details  see \citep{W07}).

For the afterglow part of the light curve we have computed values $L^*_X$ (eq. \ref{eq: lx}) at the time $T_a$, which marks the end of the plateau phase and the beginning of the last power law decay phase. We have considered the following approximation which takes into accounts the functional form, $f_a$, of the afterglow component only:

\begin{equation}
f(T_a) \approx f_a(T_a)=F_a \exp{\left ( - \frac{T_p}{T_a} \right )}  \  \  {\rm  for}  \ \ t  = T_a \\
\label{eq: fluxafterglow}
\end{equation}

\noindent
where we put the time of initial rise, $t_a=T_p$ because in most cases the afterglow component is fixed at the transition time of the prompt emission, $T_p$  (for details see \citet{W07}). Then, with applying eq-s \ref {eq: fluxafterglow} and \ref {eq: fx} in eq. \ref {eq: lx} one obtains:
\begin{equation}
L^*_X=\frac{4\pi D_L^2(z) F_X }{(1+z)^{1-\beta_{a}}}
\end{equation}
where $F_X=F_a \exp( -\frac{T_p}{T_a})$ is the observed flux at the time $T_a$. We have derived a spectral index $\beta_a$ for each GRB afterglow using the Evan's web page http://www.swift.ac.uk/xrt curves \citep{Evans2009} setting a filter time as $T_a \pm \sigma_{T_a}$; the $T_a$ values together with their errorbars, $\sigma_{T_a}$, are derived in the fitting procedure used by \citet{W07}. As mentioned above the power law spectrum, $\Phi(E) \propto E^{-(\beta_{a}+1)}$, was fitted with 
the model `phabs*phabs*pow' providing the X-ray spectral index, $\beta_a$. The first absorption component
is frozen at the Galactic column density value obtained with the NH FTOOL \footnote{ http://heasarc.gsfc.nasa.gov/lheasoft/ftools} and the second is the 'zphas' component with the redshift frozen at the value reported in the literature. For further details of the spectral fitting procedure see  \citet{Evans2009}. The lightcurves used for the analysis are the same used in \citet{Evans2009}, but binned by us in a different way.  

For some of the derived points ($L^*_{X}$, $T^*_a$) the error bars are large, indicating that the canonical lightcurve doesn't fit well the observed lightcurve. We have decided to include such cases in the analysis to treat the whole sample in a homogeneous way. Even if points with largest - a few orders of magnitude - error bars have no physical meaning, they carry information about the light curve irregularity (deviation from the considered model) or insufficient amount of observational data for precise fitting.

A choice of the Willingale model as a representation for the X-ray GRBs lightcurves allows us to use a homogeneous sample of events to study physical correlation in a statistical way. Let us point out that the fitting procedure can yield values of Ta within the gap between the end of BAT and the beginning of XRT observations, like in the case of GRB 050318. One can note that several authors fit the afterglow part of the lightcurve without modelling the prompt emission light curve. Thus, they can obtain a nearly perfect power law fits in cases where the Willingale model fitting finds a short afterglow plateau phase. 
   
To analyse how accuracy of fitting the canonical lightcurve, (eq. \ref{eq: fluxtot} and \ref{eq: fc})  to the data influences the studied correlations we use the respective logarithmic errors bars, $\sigma_{L^*_{X}}$ and $\sigma_{T^*_a}$, to formally define a fit error parameter $u \equiv \sqrt{\sigma_{L^*_{X}}^2 + \sigma_{T^*_a}^2}$, as measured in the burst rest frame. This definition is used to distinguish the canonical shaped light curves from the more irregular ones, perturbed by `secondary' flares and various non-uniformities. The symmetric error bars quoted in the paper are computed with the method of \citet{Dago05} that takes into account the hidden errors and thus gives greater error estimates than the ones obtained with the Marquardt Levemberg algorithm \citep{Marquardt}.

Our analyzed sample of 77 GRBs from the redshift range $0.08-8.26$ includes afterglows of 66 long GRBs and 11 GRBs whose nature is debated, the intermediate class (IC) between long and short GRBs, described by \cite{Norris2006} as an apparent (sub)class of bursts with a short initial pulse followed by an extended low-intensity emission phase. Our long GRB sample includes also 8 X-Ray Flashes (XRFs)(060108, 051016B, 050315, 050319 \citep{Gendre2007}, 050401, 050416A, 060512, 080330 \citep{Sakamoto2008}). XRFs  are scattered within the long GRBs distribution in Fig. \ref{fig1}, providing further support to a hypothesis that both these phenomena have the same progenitors \citep{Ioka2001}. To study physically homogeneous samples we decided here to analyze the sub-samples of 66 long GRBs (including XRFs) and of 11 IC ones separately.  

\section{The results}

\begin{figure}
\includegraphics[width=0.5\hsize,angle=-90,clip]{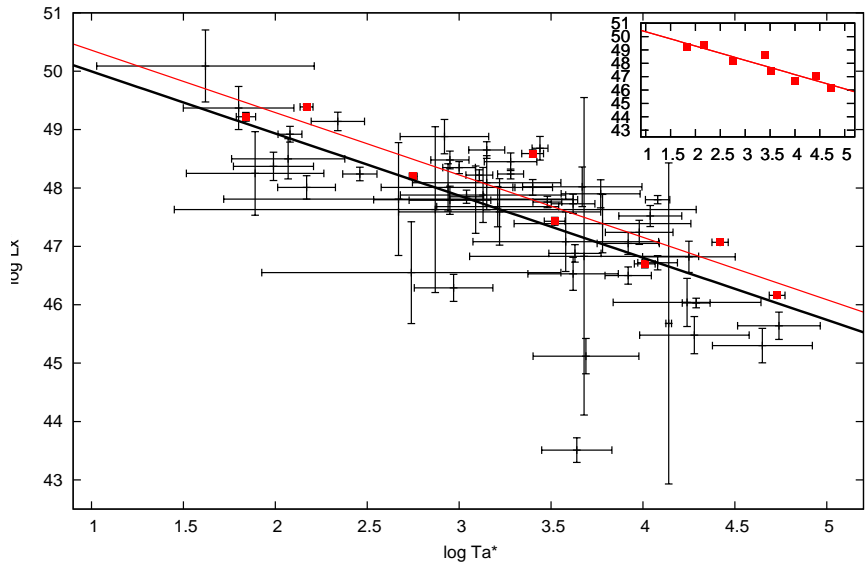}
\caption{$L^*_X$ versus $T^*_a$ distribution for the sample of 62 long afterglows with $u<4$, with the fitted correlation line in black. The upper red line, fitted to the 8 lowest error (red) points, forms approximately an upper envelope of the full distribution. The upper envelope points with the fitted line are separately presented in an inset panel. \label{fig1}}
\end{figure}

\begin{figure}
\includegraphics[width=0.5\hsize,angle=-90,clip]{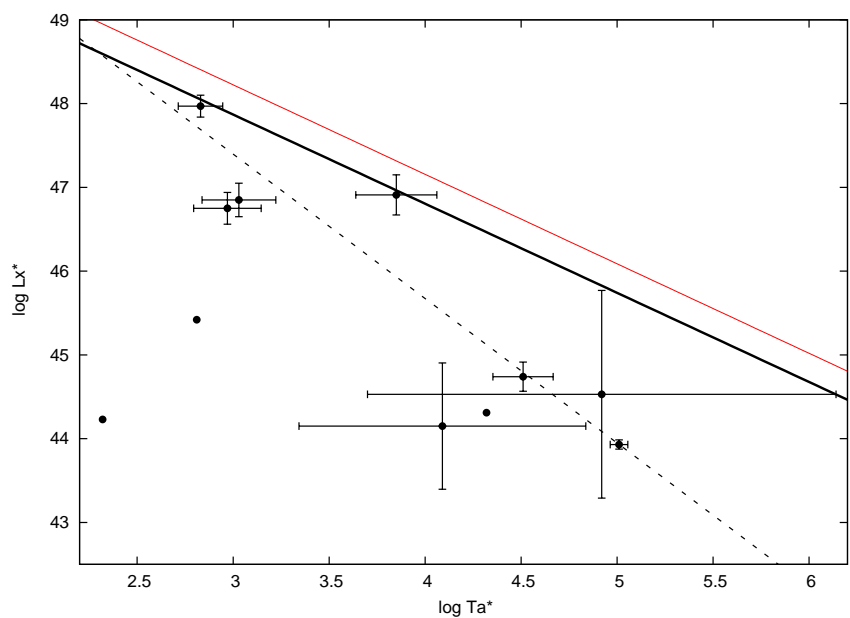}
\caption{$L^*_X$ versus $T^*_a$ distribution for the sample of 11 intermediate class GRBs. For the picture clarity the three points with very high errors bars ($u>4$) are shown without error bars. The fit dashed line is presented for the 8 points with indicated error bars, for $u<4$. Additionally, both fit lines for long GRBs from the Fig. \ref{fig1} are provided for a reference. \label{fig2}}
\end{figure}

\begin{figure}
\includegraphics[width=0.5\hsize,angle=-90,clip]{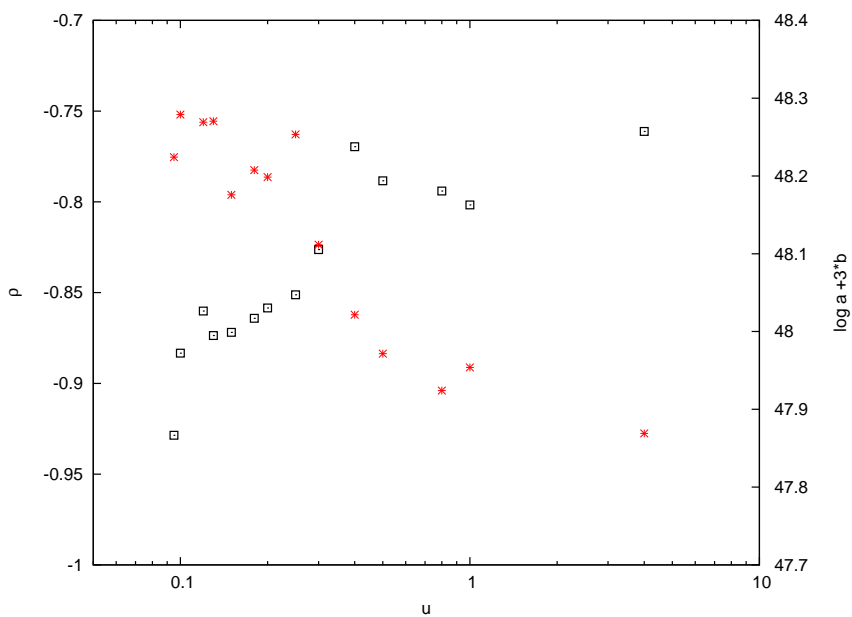}
\caption{The left vertical axis: correlation coefficients $\rho( \log L^*_X, \log T^*_a)$ versus the error parameter $u$ for the long GRBs are presented with black squares. The right vertical axis: normalizations of the fitted correlation lines at $\log T^*_a = 3.0$ versus $u$ are presented with red asterisks. 
\label{fig3}}
\end{figure}

The obtained `$L^*_X$ versus $T^*_a$' distributions  for long GRBs (Fig. \ref{fig1}; \footnote{see the data table for all long and IC GRBs at http://www.oa.uj.edu.pl/M.Dainotti/GRB2010}) and for a smaller sample of IC GRBs (Fig. \ref{fig2}) clearly demonstrate  existence of significant LT correlations, characterized in this paper by the Spearman correlation coefficient, $\rho$, a non-parametric measure of statistical dependence between two variables \citep{Spearman}. From a visual inspection of Fig. \ref{fig1} and the analysis discussed later on Fig. \ref{fig3} one can note that the lowest error events concentrate in the upper part of the distribution, forming a highly correlated subsample of the full distribution. To visualize this effect we decided to arbitrarily select 8 points with smallest errors to define our limiting the {\it upper envelope} subsample, $u<0.095$, see the inset panel in Fig. \ref{fig1}. 

For the full sample of 66 long GRBs one obtains $\rho_{LT} \equiv \rho( \log L^*_X, \log T^*_a) = -0.68$ and a probability of occurring of such correlation by chance within the uncorrelated sample $P=7.60 \times 10^{-9}$  (cf. \citet{Bevington}). If we remove a few large error points by imposing a constraint $u<4$ we have a limited sample of 62 long GRBs presented in Fig.\ref{fig1}, with $\rho_{LT} = -0.76$, $P=1.85 \times 10^{-11}$ and the fitted correlation line parameters\footnote{One may note that the presented fitted slope is different from the slope range quoted in \citet{DCC}, because in the previous paper in an attempt to reduce the intrinsic scatter in the correlation, the authors limited the sample to the GRBs with $\log L^*_X >10^{45}$ $erg/s$ and time parameter $log [T^*_a < 5$, with a possible resulting bias in the fitted correlation parameters.} $\log a = 51.06\pm 1.02$ and $b = -1.06_{-0.28}^{+0.27}$, while for the {\it upper envelope} sample we obtain, respectively, $\rho_{LT} = -0.93$, $P = 1.7 \times 10^{-2}$, $\log a = 51.39\pm 0.90$ and $b = -1.05_{-0.20}^{+0.19}$.

\begin{figure}
\includegraphics[width=0.5\hsize,angle=-90,clip]{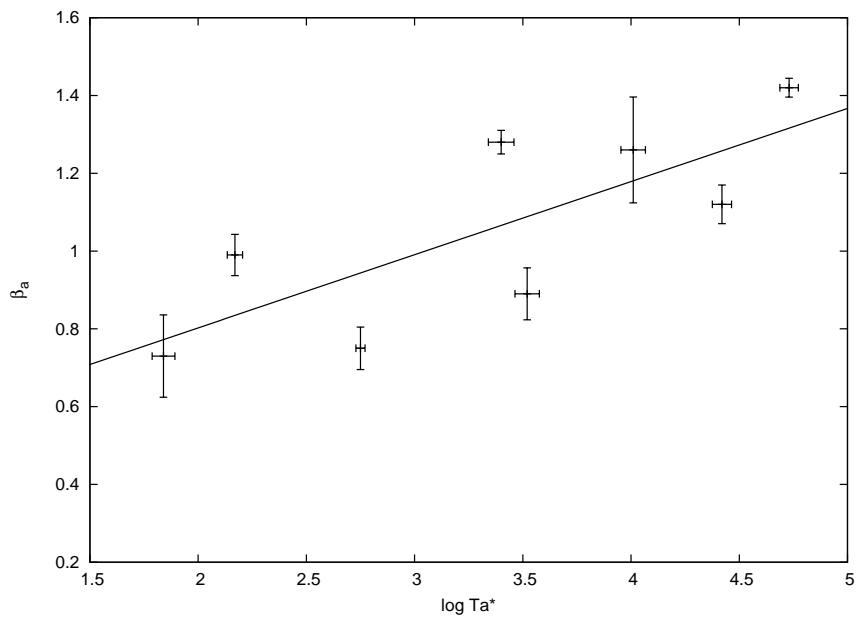}
\caption{$T^*_a$ versus the spectral index $\beta_a$ for the upper envelope sample, $u<0.095$. 
\label{fig4new}}  
\end{figure}

\begin{figure}
\includegraphics[width=0.5\hsize,angle=270,clip]{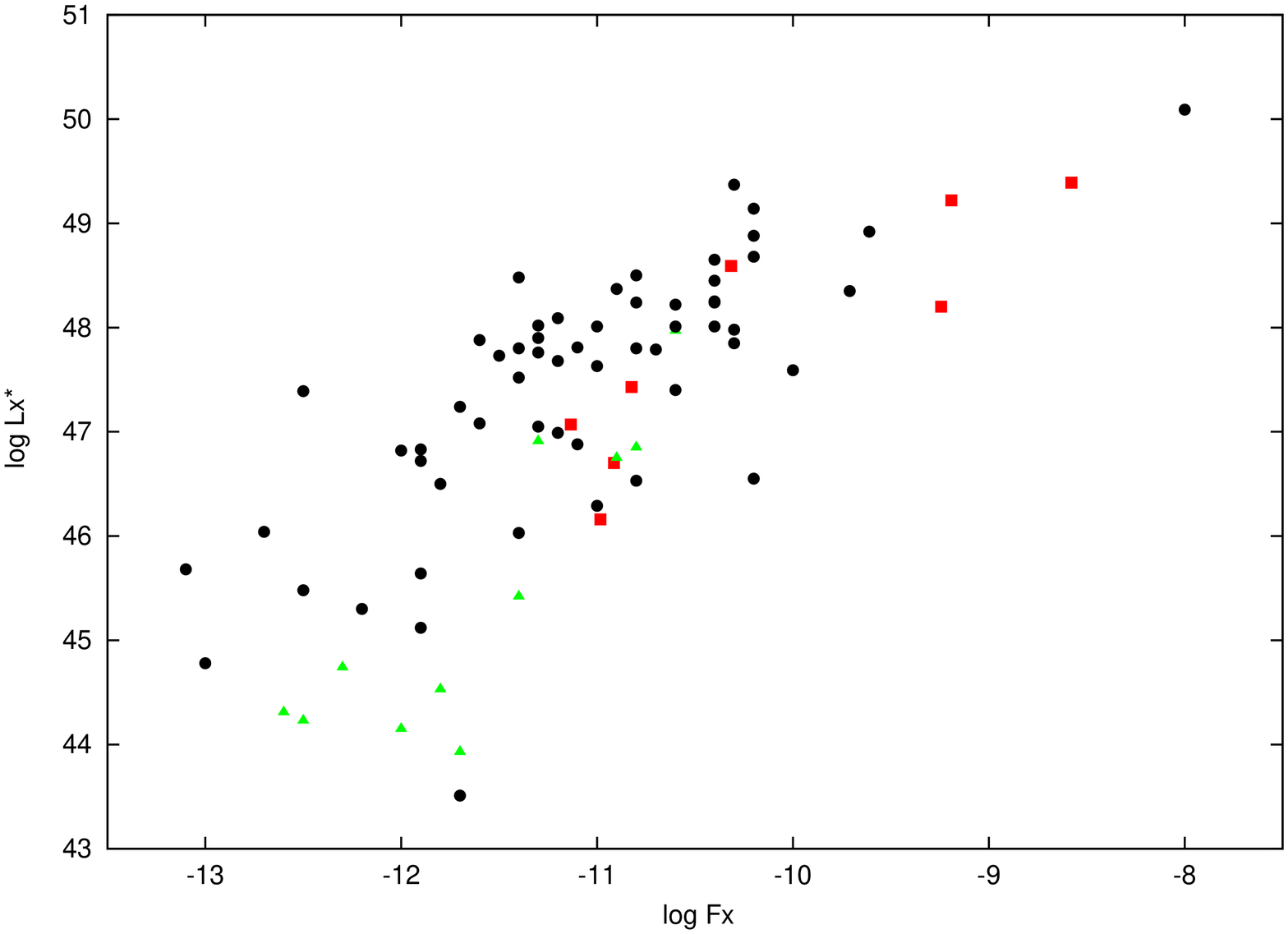}
\caption{$L^*_X$ versus the measured energy fluxes $F_X$ for the whole GRB sample. The 8 our {\it upper envelope} points are presented as red squares, while the IC GRBs are green triangles.\label{fig4}}  
\end{figure}

In Fig. \ref{fig2} we present data for all IC GRBs. One may note that their afterglows are characterized with the values of $T^*_a$ in the upper times range of the long GRBs. The IC GRBs in the $u<4$ sample -- 050724, 051221A, 060614, 060502, 070810, 070809, 070714 \citep{Norris2009}, 060912A \citep{Levan2007}  -- follow a similar LT relation as the long ones. A formally computed correlation coefficient for these 8 GRBs is $\rho_{LT} = -0.66$. This result suggests the existence of an another steeper LT correlation for IC GRBs as the one obtained for long GRBs, with different fitted parameters $\log a = 52.57\pm 1.04$ and $b = -1.72_{-0.21}^{+0.22}$. The plotted fit line is below the analogous $u<4$ sample of long GRBs showing that the IC GRBs and the normal long ones behave differently, but a limited number of available IC GRBs inhibits us to make a strong statement in that matter. 

To study the fit error systematic of GRB afterglows we show below, in Fig \ref {fig3}, how the limiting upper value for $u$ in the analyzed sample, i.e. how selecting the afterglows with increasing precision of $L^*_X$ and $T^*_a$ fits, influences the LT correlation. We present changes of the $\rho_{LT}$  converging -- with decreasing $u$ -- toward a nearly linear LT relation, as observed for our {\it upper envelope} sample. On the figure, e.g., we have 62 long GRBs for $u=4$, 33 GRBs for $u=0.3$, 19 GRBs for $u=0.15$, 13 GRBs for $u=0.12$ and 8 GRBs left for our limiting $u=0.09
5$ . A presented accompanying systematic shift upwards of the fitted correlation -- as measured in the middle of the distribution as $\log a - 3.0 \cdot b$ (the fitted correlation line at an arbitrarily selected $\log T^*_a = 3.0$) -- with decreasing $u$, proves that the limiting $u \ll 1$ subsample forms the upper part, the brightest afterglows in the LT distribution. This regular trend allows us to conclude that the subclass of all long GRBs with `canonical' afterglows forms a well defined physical class of sources exhibiting high correlation of their afterglow parameters. Presence of GRBs with light curves deviating from the \citet{W07} model increases the scatter in the L-T distribution, with larger error points distributed preferentially below the small error ones. 

We expect that having a tight LT correlation for canonical light curves, suggesting a precise physical scaling to exist between GRBs with different luminosities, should be accompanied with a regular change of the fitted spectral indices $\beta_a$, if any such changes occur besides a scatter due to measurement errors. In fact we find the existence of $\beta_a$-$\log T^*_a$ (and, of course, $\beta_a$-$\log L^*_X$) correlation, see Fig. \ref{fig4new}, for a sample of small $u<0.095$ points, where $\rho_{\beta_a T} \equiv \rho (\beta_a\, \log T^*_a) = 0.74$, but a non-negligible  probability to occur by chance from an uncorrelated distribution, $P = 1.0 \times 10^{-1}$. It suggests that the $\beta_a$ tends to increase for larger $T^*_a$: the fitted relation reads $\beta_a = 0.43 + 0.19\times \log T^*_a$. For large limiting $u=4$ this correlation becomes weak with $\rho_{\beta T} \sim 0.16$. A warning should be considered. Since we determine the value of $\beta_a$ with a filter time $T_a \pm \sigma_{T_a}$ the measurement systematics could be introduced to the data if $\beta$ varies with time in real objects. To evaluate such possibility we computed correlations $\rho (\beta_a,\, \log T_a$) using the observer measured times $T_a$, without a $z$-correction, the ones used in real $\beta_a$ measurements. However -- having a large scatter of $\beta_a$ values at all $T_a$ -- we do not observe any increase or systematic change of correlations as compared to the ones using $T^*_a$ times. This test neither excludes nor supports systematics in the data, leaving the possibility that the observed effect could be real.

To preliminary verify if any redshift systematics exists in the GRB afterglow distribution we have also studied the correlation $\rho_{Lz}\equiv \rho (\log L^*_X, z)$. We note a positive coupling, $\rho_{Lz}=0.53$, for the full long GRBs sample, but - contrary to the $\rho_{\beta T}$ distribution discussed above - the $\rho_{Lz}$ values don't have an increasing trend, fluctuating between $0.36$ and $0.61$ for smaller $u$ subsamples. For our {\it upper envelope} sample $\rho_{Lz}=0.55$, but it is  accompanied with a large random scatter of  $(\log L^*_X,z)$ points. These features does not support a clear significant redshift evolution of the GRB afterglow luminosity distribution, but the issue should be studied in more detail (Dainotti et al., in preparation). Let us also note that our limiting {\it upper envelope} subsample includes GRBs with redshifts reaching the maximum value of `only' $2.75$, while the most distant GRB with $z=8.26$ disappears from the analyzed sample after decreasing $u$ below 0.25. 

On the other hand to verify if the observed effect of higher LT correlation in the {\it upper envelope} is not a systematic effect due to higher photon statistics in brightest afterglows we compared distribution of the {\em observed} afterglow fluxes, $F_X$, for the {\it upper envelope} sample with respect to the full analyzed GRBs sample, including also the IC GRBs (Fig. \ref{fig4}). We find that the {\it upper envelope} observed flux values span more than 3 orders of magnitude and are mixed in this range with other points, in the upper part of the flux distribution. The IC GRBs have the same behavior as the long GRBs, but they are on average less luminous. Thus, the $L^*_X(T^*_a)$ and $T^*_a$ small errors result preferentially from the smooth light curve shapes, allowing precise fitting to the considered `canonical' shape, not due to a higher observed photon statistics. We note in the considered distribution that the {\it upper envelope} points are preferentially in the its lower part, with the observed systematics resulting from the LT anticorrelation (for these GRBs fluxes are measured at larger fitted $T^*_a$), but possibly influenced also by the mentioned weak $L-z$ correlation in the sample.

\section{Summary}

In the presented analysis we discovered that the afterglow light curves which are smooth and well fitted by the considered canonical model, belong to most luminous GRBs forming the well correlated upper part of the ($\log T^*_a$, $\log L^*_X$) distribution. The GRB cases with appearing flares or non-uniformities of the light curves exhibit a trend to have lower luminosities for any given $T^*_a$. We also noted the possible correlation of the X-ray spectral index $\beta_a$ and the time $T^*_a$, which, together with the LT correlation, provide new constraints for the physical model of the GRB explosion mechanism. Let us also note that the revealed tight LT correlation, if supported with larger statistics, could be a basis for a new independent cosmological test \citep{Cardone2010}.

An LT correlation for the independently analyzed (small) sample of IC GRBs is also revealed. It is different from the long GRBs, with a higher inclination of the fitted correlation lines and its luminosity normalization below the one for long GRBs.
It provides a new argument for a separate physical reality of the postulated IC GRB sub-class. Any future attempt to study relations between various GRB properties should involve, in our opinion, a separation of the IC GRBs from the long ones, to limit analysis to physically homogeneous sub-samples (like our {\it upper envelope} one). A simple increasing the studied GRB sample with a mixed content may smooth out any existing relation.

We do not intend to discuss here consequences of these findings for GRB physical models. Let us simply note that the LT relation is predicted by the models of \citet{Cannizzo09} and \citet{Ghisellini2009}, proposed for the physical GRB evolution in the time $T_a$. The \citet{Cannizzo09} model predicts a steeper correlation slope ($3/2$) than the observed one ($\approx 1$), which on the other hand is in a good agreement with the model of \citet{Yamazaki09}.

\section{Acknowledgments}
This work made use of data supplied by the UK Swift Science Data Centre at the University of Leicester. MGD and MO are grateful for the support from Polish MNiSW through the grant N N203 380336. MGD is also grateful for the support from Angelo Della Riccia Foundation and VFC for the support by University of Torino and Regione Piemonte and by INFN for the project PD51. We are particularly grateful for the anonymous referee, whose criticism helped us to improve the paper in a substantial degree.


\begin{thebibliography}{99}

\bibitem[Amati et al.(2008)]{Amati08}
Amati, L., et al., Mon. Not. R. Astron. Soc. 391, 577-584 (2008) 

\bibitem[Bevington \& Robinson (2003)]{Bevington}
Bevington, P. R., \& Robinson, D. K. 2003, Data Reduction and Error Analysis
for the Physical Sciences (3rd ed.; New York: McGraw-Hill)

\bibitem[Butler et al.(2007)]{Butler2007}
Butler, N. R., Kocevski, D., Bloom, J. S., et al. Astrophys. J., 671 656-677, (2007)

\bibitem[Butler et al.(2009)]{Butler2009}
Butler, N. R., Bloom, J. S., Poznanski, D. arXiv:0910.3341 (2009) 

\bibitem[Butler et al.(2009)]{Butler2010}
Butler, N. R., Bloom, J. S., Poznanski, D. Astrophys. J., 711 495B, (2010)

\bibitem[Bloom et al.(2001)]{B01}
Bloom, J.S., Frail, D.A., Sari, R. Astron. J. 121, 2879-2888 (2001)

\bibitem[Cannizzo \& Gehrels (2009)]{Cannizzo09}
Cannizzo,J. K. \& Gehrels, N.  Astrophys. J. 700, 1047-1058 (2009)

\bibitem[Cabrera et al.(2009)]{Cabrera2007}
Cabrera J. I., Firmani, C., Ghisellini, et al. Mon. Not. R. Astron. Soc. 382, 342 (2007)

\bibitem[Cardone et al.(2009)]{Cardone2010}
Cardone, V.F, Capozziello, S. \& Dainotti, M.G. Mon. Not. R. Astron. Soc. 400, 775 (2009) 

\bibitem[Cardone et al.(2009)]{Cardone09}
Cardone, V.F, Dainotti, M.G., Capozziello, S. \& Willingale, R. Mon. Not. R. Astron. Soc. 400, 775 (2009) 

\bibitem[Dainotti et al.(2008)]{DCC}
Dainotti, M.G., Cardone, V.F., Capozziello, S. Mon. Not. R. Astron. Soc. 391, L79-L83 (2008)

\bibitem[D'Agostini(2005)]{Dago05}
D' Agostini, G. arXiv\,:\,physics/0511182

\bibitem[Evans et al.(2009)]{Evans2009}
Evans,P. et al. Mon. Not. R. Astron. Soc. 397, 1177-1201 (2009) 

\bibitem[Fenimore \& Ramirez\,-\,Ruiz(2000)]{FRR00}
Fenimore, E.E., Ramirez\,-\,Ruiz, E. 2000, Astrophys. J. 539, 712-717 (2000)

\bibitem[Firmani et al.(2006a)]{fir06}
Firmani, C., Ghisellini, G., Avila-Reese, V., \& Ghirlanda, G. Mon. Not. R. Astron. Soc. 370, 185 (2006)

\bibitem[Gendre et al.(2007)]{Gendre2007}
Gendre, B, Galli, A.  Piro, L. \& A\&A 465, L13�L16 (2007)

\bibitem[Ghirlanda et al.(2006)]{Ghirlanda06}
Ghirlanda G., Ghisellini G. \& Firmani C.,  New J. of Phys. 8, 123 (2006)

\bibitem[Ghirlanda et al.(2004)]{G04}
Ghirlanda, G., Ghisellini, G., Lazzati, D. Astrophys. J. 616, 331 (2004)

\bibitem[Ghisellini et al.(2009)]{Ghisellini2009}
Ghisellini G., Nardini, M., Ghirlanda G., Celotti, A., Mon. Not. R. Astron.  393 253-271 (2009)

\bibitem[Hu \& Sugiyama(1996)]{HS96}
Hu, W., Sugiyama, N. ApJ, 471, 542 (1996)

\bibitem[Marquardt(1963)]{Marquardt}
Marquardt, D. SIAM J. on Applied Phys.  11, 431- 441 (1963)

\bibitem[Yamazaki(2009)]{Yamazaki09}
Yamazaki,R. (2009), Astrophys. J.  690, L118-L121 (2009)

\bibitem[Izzo et al.(2009)]{Izzo09}
Izzo, L. Capozziello, S., Covone, G., Capaccioli, M. 2009, accepted on Astron. \& Astrophys.  2009arXiv0910.1678I

\bibitem[Ioka \& Nakamura(2001)]{Ioka2001}
Ioka, K. \& Nakamura, T. \aap 554 L163-L167 (2001)

\bibitem[Levan et al.(2007)]{Levan2007}
Levan, A.J., Jakobsson, Hurkett, MNRAS 378 1439-1446, (2007) 

\bibitem[Liang \& Zhang(2006)]{LZ06}
Liang, E.W., Zhang, B. Mon. Not. R. Astron. Soc. 369, L37-L41 (2006)

\bibitem[Liang \& Zhang(2005)]{liza05}
Liang, E., \& Zhang, B. Astrophys. J. 633, 611-623 (2005)

\bibitem[Liang et al.(2009)]{LWZ09}
Liang,N., Wu, P. \& Zhang, S.N arXiv:0911.5644v2, (2009)

\bibitem[Norris et al.(2000)]{N000}
Norris, J.P., Marani, G.F., Bonnell, J.T. Astrophys. J. 534, 248-257 (2000)

\bibitem[Norris et al.(2006)]{Norris2006}
Norris, J.P. \& Bonnell, J.T. \aap, 643, 266-275, (2006)

\bibitem[Norris et al.(2006)]{Norris2009}
Norris, J.P. \& Bonnell, J.T. arXiv:0910.2456,(2009)

\bibitem[Nousek et al.(2006)]{Nousek06}
Nousek, J.A, Kouveliotou, C., Grupe, D. et al. \aap 642, 389 (2006)

\bibitem[Riess et al.(2007)]{Riess07}
Riess, A.G et al. , Astrophys. J. 659, 98-121 (2007)

\bibitem[Salvaterra et al.(2009)]{Salvaterra2009}
Salvaterra,R., et al. accepted to Nature, arxiv:0906.1578 (2009)

\bibitem[Sakamoto (2008)]{Sakamoto2008}
Sakamoto T., AIP Conference Proceedings, 1065, 9-12 (2008)

\bibitem[Shahmoradi \& Nemiroff (2009)]{Shahmoradi09}
Shahmoradi, A. \& Nemiroff R. J. arXiv0904.1464S (2009)

\bibitem[Schaefer(2007)]{S07}
Schaefer, B.E. Astrophys. J.  660, 16-46 (2007)

\bibitem[Spearman (1904)]{Spearman}
Spearman, C. The American Journal of Psychology, 15, 72 (1904) 

\bibitem[Qi et al. (2009)]{Qi09}
Qi, S., Lu, T., Wang, F.-Y., MNRAS 398L, 78-82 (2009)

\bibitem[Tanvir et al. (2009)]{Tanvir}
Tanvir, N. R. at al. Nature 461, 7268 (2009)

\bibitem[Willingale et al.(2007)]{W07}
Willingale, R.W. et al., Astrophys. J.  662, 1093-110 (2007)

\bibitem[Yu et al.(2009)]{Yu09}
Yu, B., Qi, S., Lu, T. Astrophys. J. Letters 705 L15-L19 (2009) 

\bibitem[Zhang et al.(2006)]{Zhang06}
Zhang B., Fan Y.Z. et al., Astrophys. J.  642, 354-370 (2006).

\end{thebibliography}
\end{document}